\documentclass[12pt]{article}
\usepackage{graphicx}

\newcommand\pubnumber{}
\newcommand\pubdate{\today}

\def\Title#1{\begin{center} {\Large #1 } \end{center}}
\def\Author#1{\begin{center}{ \sc #1} \end{center}}
\def\Address#1{\begin{center}{ \it #1} \end{center}}

\newcommand\pubblock{\rightline{\begin{tabular}{l} \pubnumber\\
         \pubdate  \end{tabular}}}
\newenvironment{Abstract}{\begin{quotation}  }{\end{quotation}}
\newenvironment{Presented}{\begin{quotation} \begin{center} 
             PRESENTED AT\end{center}\bigskip 
      \begin{center}\begin{large}}{\end{large}\end{center} \end{quotation}}

\def\beq{\begin{equation}}
\def\eeq#1{\label{#1}\end{equation}}
\def\eeqn{\end{equation}}

\def\beqa{\begin{eqnarray}}
\def\eeqa#1{\label{#1}\end{eqnarray}}
\def\eeqan{\end{eqnarray}}

\let\bar=\overbar

\def\Dslash{\not{\hbox{\kern-4pt $D$}}}
\def\dslash{\not{\hbox{\kern-2pt $\del$}}}

\def\msb{{\bar{\ssstyle M \kern -1pt S}}}

\usepackage{graphicx,xcolor}
\usepackage{mathtools,amssymb}
\usepackage{multirow}
\usepackage{braket}
\usepackage{relsize}
\usepackage{booktabs}
\usepackage{tikz,multicol}

\newcommand{\pmatr}[1]{\begin{pmatrix} #1 \end{pmatrix}}
\newcommand{\simlt}{~\mbox{\smaller\(\lesssim\)}~}
\newcommand{\simgt}{~\mbox{\smaller\(\gtrsim\)}~}

\begin{document}
\begin{titlepage}
\pubblock

\vfill
\Title{Towards a complete $ \Delta(27)\times SO(10) $ SUSY GUT}
\vfill
\Author{Fredrik Bj\"orkeroth}
\Address{School of Physics and Astronomy, University of Southampton, SO17 1BJ Southampton, UK}
\vfill
\begin{Abstract}
I discuss a renormalisable model based on $\Delta(27)$ family symmetry with an $SO(10)$ grand unified theory (GUT) with spontaneous geometrical $CP$ violation.
The symmetries are broken close to the GUT breaking scale, yielding the minimal supersymmetric standard model.
Low-scale Yukawa structure is dictated by the coupling of matter to $ \Delta(27) $ antitriplets $ \bar{\phi} $ whose vacuum expectation values are aligned in the CSD3 directions by the superpotential.
Light physical Majorana neutrinos masses emerge from the seesaw mechanism within $SO(10)$. 
The model predicts a normal neutrino mass hierarchy with the best-fit lightest neutrino mass $ m_1 \sim 0.3 $ meV, $CP$-violating oscillation phase $\delta^l\approx 280^{\circ}$ and the remaining neutrino parameters all within $1 \sigma$ of their best-fit experimental values.
\end{Abstract}
\vfill
\begin{Presented}
NuPhys2016, Prospects in Neutrino Physics \\
Barbican Centre, London, UK,  December 12--14, 2016
\end{Presented}
\vfill
\end{titlepage}
\section{Introduction}
It is well established that the Standard Model (SM) remains incomplete while it fails to explain why neutrinos have mass. Small Dirac masses may be added by hand, but this gives no insight into the Yukawa couplings of fermions to Higgs (where a majority of free parameters in the SM originate), or the extreme hierarchies in the fermion mass spectrum, ranging from neutrino masses of $ \mathcal{O}(\mathrm{meV}) $ to a top mass of $ \mathcal{O}(100) $ GeV. 
Understanding this, and flavour mixing among quarks and leptons, constitutes the flavour puzzle.
Other open problems unanswered by the SM include the sources of $CP$ violation (CPV), as well as the origin of three distinct gauge forces, and why they appear to be equal at very high energy scales.

An approach to solving these puzzles is to combine a Grand Unified Theory (GUT) with a family symmetry which controls the structure of the Yukawa couplings. 
In the highly attractive class of models based on $SO(10)$
\cite{Fritzsch:1974nn}
, three right-handed neutrinos are predicted and neutrino mass is therefore inevitable via the seesaw mechanism.

In this paper I summarise a recently proposed model \cite{Bjorkeroth:2015uou}, renormalisable at the GUT scale, capable of addressing all the above problems, based on  $\Delta(27) \times SO(10)$.

\section{Features of the model}

All SM fermions (and their superpartners) are gathered in a single superfield $ \Psi $, an $ SO(10) $ spinor that is a triplet under $ \Delta(27) $, and couples to (gauge singlet) antitriplet flavons $ \bar{\phi} $. The vacuum expectation values (VEVs) of these flavons are aligned in particular directions in flavour space, known as the CSD3 alignment, which dictate the Yukawa structure at the low scale.

Large lepton mixing is accounted for by the seesaw mechanism 
\cite{Minkowski:1977sc, Ramond:1979py} 
with constrained sequential dominance (CSD) \cite{King:1998jw}. 
The basic goal of the flavour sector in these models is to couple matter to flavons $ \bar{\phi}_{\rm atm} $, $ \bar{\phi}_{\rm sol} $ and $ \bar{\phi}_{\rm dec} $, whose VEVs are aligned in the CSD3 direction \cite{King:2013iva}, i.e. where
\begin{equation}
	\bar{\phi}_\mathrm{atm} = v_\mathrm{atm} \pmatr{0\\1\\1} , \qquad \bar{\phi}_\mathrm{sol} = v_\mathrm{sol} \pmatr{1\\3\\1} , \qquad \bar{\phi}_\mathrm{dec} = v_\mathrm{dec} \pmatr{0\\0\\1},
	\label{eq:csd3alignments}
\end{equation}
which has been previously shown to be a very promising and predictive setup for understanding lepton masses \cite{Bjorkeroth:2014vha}.

Since $ SO(10) $ constrains the Dirac couplings of leptons and quarks to be equal (or nearly so), it is non-trivial that the successful scheme in the lepton sector will translate to success in the quark sector. Remarkably we find that we can attain good fits to data for quark and lepton masses, mixings and phases. 

Interactions between the flavons and additional non-trivial singlets of $\Delta(27)$ give rise to CPV phases, which fixes all phases in the lepton mass matrices, and leads to a novel form of spontaneous geometrical CPV (for a partial list of available literature on geometrical CPV, see Ref.~6 in \cite{Bjorkeroth:2015uou}).

The model has many other attractive features, including the use of only the lower dimensional ``named'' representations of $SO(10)$, i.e. the singlet, fundamental, spinor or adjoint representations.
$SO(10)$ is broken via $SU(5)$ with doublet-triplet (DT) splitting achieved by a version of the Dimopoulos-Wilczek (DW) mechanism \cite{DW}.
We find that the MSSM $ \mu $ term can naturally be $ \mathcal{O}(\mathrm{TeV}) $, and confirm that proton decay is highly suppressed.

The family symmetries are broken close to the GUT breaking scale to yield the minimal supersymmetric standard model (MSSM).
The model is realistic in the sense that it provides a successful (and natural) description of the quark and lepton (including neutrino) mass and mixing spectra.

\section{Field content and mass matrices}
The most important superfields in the model \cite{Bjorkeroth:2015uou} are given in table \ref{tab:fields}. It includes the MSSM matter superfield $ \Psi $, various Higgs fields that break $ SO(10) $, $ SU(5) $ and electroweak gauge symmetries, and flavons $ \bar{\phi}_i $ that produce the phenomenologically successful form of the quark and lepton mass matrices. 
It also includes the singlet $ \xi $, which acquires a VEV below the GUT scale, $ \braket{\xi} \! \simlt 0.1 M_\mathrm{GUT} $ and ultimately controls the hierarchies present in the model.
$ CP $ is conserved at the high scale and spontaneously broken by flavons, while GUT and flavour singlets $ Z, Z^{\prime\prime} $ break $ \mathbb{Z}_4^R $ $R$-symmetry to regular $R$-parity
\cite{Lee:2011dya}.

\begin{table}[ht]
\centering
\begin{tabular}{@{}l*{15}{l}}
\toprule
	Field	& $\Delta(27)$ & $SO(10)$ & $\mathbb{Z}_4^R$ & Role\\[0.75ex]
\midrule
	$ \Psi $ & 3 & 16 & 1 & Contains SM fermions \\
	$ H_{10}^{u,d} $ & 1 & 10 & 0 & Break electroweak symmetry \\
	$H_{16,\bar{16}}$ & 1 & $16, \bar{16}$ & 0 & Break $ SO(10) $ \\
	$ H_{45} $ & 1 & 45 & 0 & Breaks $SU(5)$ \\
	$ H_{DW} $ & 1 & 45 & 2 & Gives DT splitting via DW mechanism \\
	$\bar{\phi}_{i} $ & $ \bar{3} $ & 1 & 0 & Produces CSD3 mass matrices \\ 
	$ \xi $ & 1 & 1 & 0 & Gives mass hierarchies, $ \mu $ term \\
	$ Z, Z^{\prime\prime} $ & 1 & 1 & 2 & Break $ \mathbb{Z}_4^R \rightarrow \mathbb{Z}_2^R $ ($R$-parity) \\
	$ A_i $ & 3 & 1 & 2 & Aligns triplet flavons $\bar{\phi}_{i} $ \\
	$ O_{ij} $ & $1_{ij}$ & 1 & 2 & Aligns triplet flavons $\bar{\phi}_{i} $ \\
\bottomrule
\end{tabular}
\caption{Core superfields of the model and their representations under the symmetries.}
\label{tab:fields}
\end{table}

The flavour structure for quark and lepton matrices is determined by the CSD3 framework, where the fermions couple to flavons that acquire VEVs like in Eq.~\ref{eq:csd3alignments}. 
More precisely, its implementation can be understood as follows: at the renormalisable level, $ \Psi $, $ \bar{\phi}_i $ and $ H $ couple to $ SO(10) $ spinor superfields $ \chi_i $, which act as messengers. Integrating these out produces Yukawa-like terms 
\begin{equation}
	\lambda \Psi \Psi H \bar{\phi}_i \bar{\phi}_j \frac{\xi^n}{M_\chi^{n+2}} \,,
\end{equation}
where $ \lambda $ are $ \mathcal{O}(1) $ couplings, $ n $ are integer powers, and $ M_\chi \simgt M_\mathrm{GUT} $. 
The presence of different such terms with different powers of $ \xi $ establishes a hierarchy between elements of the fermion mass matrices, and consequently a natural explanation for the large range of fermion masses.
$ \xi $ is also responsible for generating a $ \mu $ term that can be $ \mathcal{O}(\mathrm{TeV}) $. After $ SO(10) $ is broken, all but two of the Higgs doublets and all triplets contained in $ H^{u,d}_{10} $ and $ H_{16,\bar{16}} $ must be very heavy, while exactly two doublets, $ H_{u,d} $, are light. Constructing the effective mass matrix for Higgs doublets, it has one light eigenvalue with $ \mu \sim \braket{\xi}^8\! /M_\mathrm{GUT}^7 \ll M_\mathrm{GUT} $.

Guided by $ SO(10) $ unification, ultimately all mass matrices take the same generic form:
\begin{equation}
	m^f = m^f_1 \pmatr{0&0&0\\0&1&1\\0&1&1} + m^f_2 \pmatr{1&3&1\\3&9&3\\1&3&1} + m^f_3 \pmatr{0&0&0\\0&0&0\\0&0&1},
\label{eq:mf}
\end{equation}
where $ m^f_i $ are complex numbers. 
An exception is the term $ \Psi \Psi H^u_{10} \xi \bar{\phi}_\mathrm{sol}\bar{\phi}_\mathrm{dec} $, which is allowed by the symmetries and messengers, giving an additional contribution to $ m^u $ like
\begin{equation}
	m_{4}^u \pmatr{0&0&1\\0&0&3\\1&3&2}.
\end{equation}
It does not affect neutrino physics. As the matrix in Eq.~\ref{eq:mf} is a sum over rank-1 matrices, we find that also the neutrino mass matrix after seesaw always retains this form. A proof is given in \cite{Bjorkeroth:2016lzs}, which also discusses leptogenesis in this model.

This matrix structure can successfully accommodate all quark and lepton masses and mixing angles, and is predictive in the lepton sector. It predicts a Normal Hierarchy with best fit parameters
\begin{alignat}{6}
	\theta_{12} &= 33.1^\circ , & \qquad 
	\delta_\mathrm{CP} &= 280^\circ ,& \qquad
	m_1 &= 0.32 {\rm ~meV},
	\nonumber \\
	\theta_{13} &= 8.55^\circ , & 
	\alpha_{21} &= 264^\circ ,&
	m_2 &= 8.64 {\rm ~meV},
	\\
	\theta_{23} &= 40.8^\circ ,&
	\alpha_{31} &= 323^\circ ,&
	m_3 &= 49.7 {\rm ~meV}.
	\nonumber
\end{alignat}
Once lepton masses are fitted, the PMNS matrix is completely fixed with no freedom, including the Dirac phase $ \delta_\mathrm{CP} $, which is a prediction of the model and not fitted. It agrees well with the current experimental hints.

\section{Conclusion}
This is one of few flavour $SO(10)$ models with realistic fits to \emph{both} quark and lepton data, and the most complete, as it is renormalisable and accounts for gauge coupling unification, the $ \mu $ term, DT splitting, and (lack of) proton decay.
It is predictive in the lepton sector, and may be probed by precise measurements of PMNS parameters, particularly $ \delta_\mathrm{CP} $, and ruled out by observation of an Inverse Hierarchy, or non-observation of SUSY.


\begin{thebibliography}{9}

\bibitem{Fritzsch:1974nn}
  H.~Fritzsch and P.~Minkowski,
  Annals Phys.\  {\bf 93} (1975) 193.

\bibitem{Bjorkeroth:2015uou}
  Bj\"orkeroth F, de Anda F J, de Medeiros Varzielas I and King S F,
  \emph{Phys.\ Rev.\ D} {\bf 94} (2016) no.1,  016006
  [1512.00850].

\bibitem{Minkowski:1977sc}
  P.~Minkowski,
  Phys.\ Lett.\  B {\bf 67} (1977) 421;
  T. Yanagida, in Proceedings of the Workshop on Unified Theory and Baryon Number of the Universe, eds. O. Sawada and A. Sugamoto (KEK, 1979) p.95;
  M. Gell-Mann, P. Ramond and R. Slansky, in Supergravity, eds. P. van Niewwenhuizen and D. Freedman (North Holland, Amsterdam, 1979) Conf.Proc. C790927 p.315, PRINT-80-0576.
\bibitem{Ramond:1979py}
  P.~Ramond, 
  Invited talk given at Conference: C79-02-25
  (Feb 1979) p.265-280, CALT-68-709,
  hep-ph/9809459.
  
\bibitem{King:1998jw}
  King S F,
  \emph{Phys.\ Lett.\  B} {\bf 439} (1998) 350
  [hep-ph/9806440];\,
  King S F,
  \emph{Nucl.\ Phys.\  B} {\bf 562} (1999) 57
  [hep-ph/9904210];\,
  King S F,
  \emph{Nucl.\ Phys.\  B} {\bf 576} (2000) 85
  [hep-ph/9912492];\,
  King S F,
  \emph{JHEP} {\bf 0209} (2002) 011
  [hep-ph/0204360];
  King S F,
  \emph{JHEP} {\bf 0508} (2005) 105
  [hep-ph/0506297];
  Antusch S, King S F, Luhn C and Spinrath M,
  Nucl.\ Phys.\ B {\bf 856} (2012) 328
  [1108.4278].

\bibitem{King:2013iva}
  King S F,
  \emph{JHEP} {\bf 1307} (2013) 137
  [1304.6264];

\bibitem{Bjorkeroth:2014vha}
  Bj\"orkeroth F and King S F,
  \emph{J.\ Phys.\ G} {\bf 42} (2015) 12,  125002
  [1412.6996].

\bibitem{DW}
  Dimopoulos S and Wilczek F, Report No. NSF-ITP-82-07 (unpublished);
  Babu K S and Barr S M,
  \emph{Phys.\ Rev.\ D} {\bf 48} (1993) 5354
  [hep-ph/9306242);
  Barr S M and Raby S,
  \emph{Phys.\ Rev.\ Lett.\ } {\bf 79} (1997) 4748
  [hep-ph/9705366].

\bibitem{Lee:2011dya}
  Lee H M, Raby S, Ratz M, Ross G G, Schieren R, Schmidt-Hoberg K and Vaudrevange P K S,
  \emph{Nucl.\ Phys.\ B} {\bf 850} (2011) 1
  [1102.3595].

\bibitem{Bjorkeroth:2016lzs}
  F.~Bj\"orkeroth, F.~J.~de Anda, I.~de Medeiros Varzielas and S.~F.~King,
  JHEP {\bf 1701} (2017) 077
  doi:10.1007/JHEP01(2017)077
  [arXiv:1609.05837 [hep-ph]].

\end{thebibliography}
\end{document}